\documentclass[twocolumn]{jpsj3arxiv}

\usepackage{bm}
\usepackage{mathrsfs}
\usepackage{exscale}
\usepackage{color}
\usepackage{graphicx}

\newcommand{\AC}{AC}
\newcommand{\ACs}{ACs}
\newcommand{\QC}{QC}
\newcommand{\QCs}{QCs}
\newcommand{\agy}{AGY}
\newcommand{\agyI}{AGY$({\rm I})$}
\newcommand{\agyII}{AGY$({\rm II})$}
\newcommand{\aay}{AAY}
\newcommand{\agyIN}{Au$_{64.0}$Ge$_{22.0}$Yb$_{14.0}$}
\newcommand{\agyIIN}{Au$_{63.5}$Ge$_{20.5}$Yb$_{16.0}$}

\newcommand{\agyIX}{Au$_{65.3}$Ge$_{20.9}$Yb$_{13.8}$}

\title{Crystal Structure of Superconducting 1/1 Cubic Au-Ge-Yb Approximant with Tsai-type Cluster}

\author{
Kazuhiko~\text{Deguchi}$^{1}$\thanks{E-mail: deguchi@edu3.phys.nagoya-u.ac.jp}, 
Mika~\text{Nakayama}$^{1}$,
Shuya~\text{Matsukawa}$^{1}$,
Keiichiro~\text{Imura}$^{1}$,
Katsumasa~\text{Tanaka}$^{2}$,
Tsutomu~\text{Ishimasa}$^{2}$, and 
Noriaki~K.~\text{Sato}$^{1}$}

\inst{$^1$Department of Physics, Graduate School of Science, Nagoya University, Nagoya 464-8602, Japan\\
$^2$Division of Applied Physics, Graduate School of Engineering, Hokkaido University, Sapporo 060-8628, Japan}

\abst{
We report the synthesis of a single-phase sample of the superconducting crystalline approximant \agyIN\ and present a structure model refined by Rietveld analysis for X-ray diffraction data.
}

\begin{document}
\maketitle

Quasicrystals (\QCs ) are metallic alloys that possess long-range, quasi-periodic structures with diffraction symmetries forbidden to conventional crystals,
while approximant crystals (\ACs ) are alloys whose composition is close to that of \QCs\ and whose unit cell has atomic decorations similar to those of the QCs.
Recently, new types of \QC\ and \AC\ have been discovered~\cite{Ishimasa2011}: The Au-Al-Yb (\aay ) \QC\ exhibits unconventional quantum critical behavior as $T \rightarrow 0$ and the \aay\ \AC\ shows heavy Fermi liquid~\cite{Deguchi2012}, and the Au-Ge-Yb (\agy ) \ACs\ show  superconductivity below 1 K~\cite{Deguchi2014}.
For the latter system, there are two types of alloys with different concentrations, \agyIN\ [\agyI ] and \agyIIN\ [\agyII ], with superconducting transition temperatures of 0.68 and 0.36 K, respectively.
In this Short Note, we report a detailed structure analysis of the \agyI\ \AC.
For \agyII, we are unable to provide a structure model owing to the lack of a single-phase sample.

Polycrystalline samples of the optimally composition-controlled compound \agyIN\ were synthesized, and their structures were characterized by powder X-ray diffraction using Cu $K\alpha$-radiation. 
Details of the sample preparation and X-ray diffraction analysis experiments were described in Refs.~\citen{Deguchi2014} and \citen{Ishimasa2011}, respectively. 

\begin{figure}[t]
\begin{center}
\includegraphics[clip,width=0.90\columnwidth]{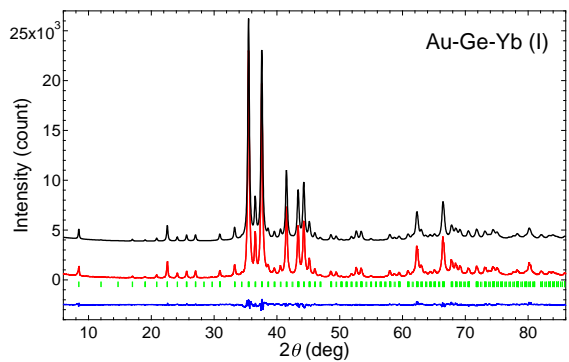}
\end{center}
\caption{(Color online) Powder X-ray diffraction pattern of \agyIN\ [\agyI ]. From top to bottom: calculated intensity, measured intensity, peak position, and difference between calculation and measurement.
}
\label{fig:X-ray}
\end{figure}
\begin{figure}[t]
\begin{center}
\includegraphics[clip,width=0.75\columnwidth]{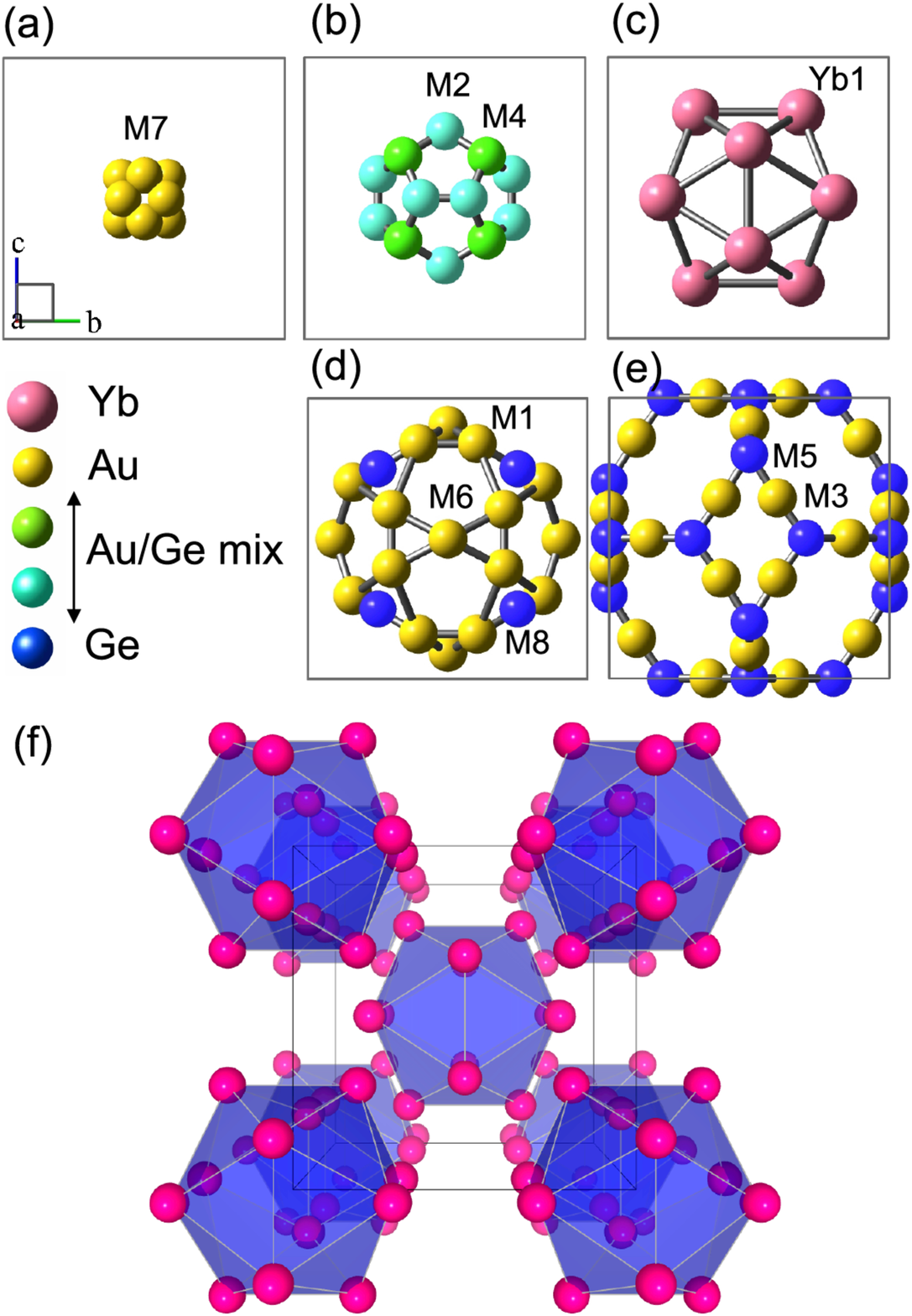}
\end{center}
\caption{(Color online) Concentric shell structure of Tsai-type cluster in the \agyI\ \AC . A square frame denotes a unit cell.  (a) Orientationally disordered tetrahedron composed of Au atoms at M7 site. (b) Second shell of dodecahedron composed of mixed Au/Ge atoms at M2 and M4 sites. (c) Third shell of icosahedron composed of Yb atom (Yb1 site). (d) Fourth shell of icosidodecahedron composed of Au atoms (M1 and M6 sites) and Ge atoms (M8 site). (e) Fifth shell of triacontahedron composed of Au atoms (M3 site) and mainly of Ge atoms (M5 site). (f) Body-centered cubic arrangement of Yb icosahedron in the projection along the [001] direction.}
\label{fig:Crystal}
\end{figure}
\begin{table*}[t]
\begin{center}
\caption{Parameters in the structure model of \agyI\ 1/1 cubic \AC .
Occupancies of sites M7 and M8 are 26.9\% and 95\%, respectively.
Note that the $B$-factor is proportional to the atomic displacement parameter $U_{\rm eq} = B/(8\pi^2)$.
}
\label{tab:X-ray}
\begin{tabular*}{\textwidth}[b]{@{\extracolsep{\fill}}lrlllll}
\hline
Site&Set&Atom&$x$&$y$&$z$&$B$ (\AA$^2$) \\
\hline
M$1$&$48h$&${\rm Au}$&$0.33941(8)$&$0.2014(1)$&$0.10469(8)$&$0.39(3)$\\
M$2$&$24g$&$0.414{\rm Au}+0.586{\rm Ge}$&$0$&$0.2381(3)$&$0.0850(2)$&$0.71(9)$\\
M$3$&$24g$&${\rm Au}$&$0$&$0.5986(2)$&$0.6431(1)$&$0.20(4)$\\
M$4$&$16f$&$0.78{\rm Au}+0.22{\rm Ge}$&$0.1489(2)$&$-$&$-$&$1.4(1)$\\
M$5$&$12e$&$0.070{\rm Au}+0.930{\rm Ge}$&$0.2012(4)$&$0$&$1/2$&$0.4(1)$\\
M$6$&$12d$&${\rm Au}$&$0.4042(2)$&$0$&$0$&$0.49(7)$\\
M$7$&$24g$&$0.269{\rm Au}$&$0$&$0.082(1)$&$0.076(1)$&$5.9(5)$\\
M$8$&$8c$&$0.95{\rm Ge}$&$1/4$&$1/4$&$1/4$&$3.0(4)$\\
Yb$1$&$24g$&${\rm Yb}$&$0$&$0.1867(2)$&$0.3046(2)$&$0.10(4)$\\
\hline
\end{tabular*}
\end{center}
\end{table*}
\begin{figure}[t]
\begin{center}
\includegraphics[clip,width=0.95\columnwidth]{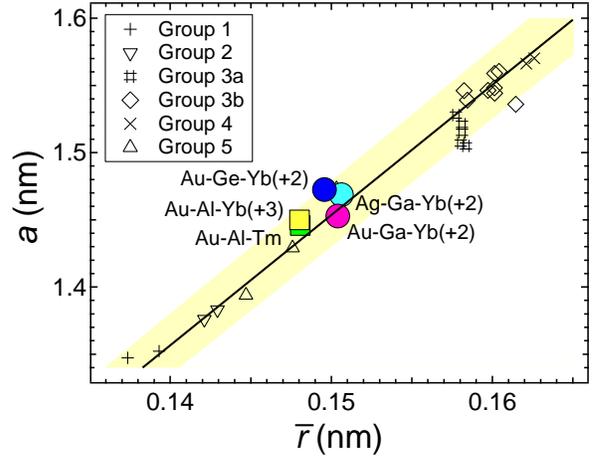}
\end{center}
\caption{(Color online) Linear relationship between lattice parameter $a$ and average atomic radius $\bar{r}$ of Tsai-type ACs. Group 1: Cu-Al-Sc and Cu-Ga-Sc. Group 2: Zn-Sc and Zn-Cu-Sc. Group 3a: Ag-In-$R$  ($R =$ rare earth) and Au-In-Ca, Group 3b: Ag-In-$R$, Cd-Y, and Cd-$R$. Group 4: Cd-Ca and Cd-Yb. Group 5: Pd-Al-Sc, Zn-Yb, and Au-Ga-Ca. See Ref.~\citen{Ishimasa2011} and references therein.}
\label{fig:AC}
\end{figure}
Figure~\ref{fig:X-ray} shows a powder X-ray diffraction pattern of \agyI, indicating the body-centered cubic structure (space group $Im\bar{3}$) with a lattice parameter $a = 1.4724(2)$ nm.
The Rietveld analysis successfully converged with $R$-factors, $R_{\rm wp} = 5.41$\%, $R_{\rm I} = 0.95$\%, and $R_{\rm F} = 0.59$\% (for the structure parameters, see Table~\ref{tab:X-ray}), 
assuming that the cluster center is occupied by an orientationally disordered tetrahedron~\cite{Lin2010}.
This suggests that the alloy has a similar structure to the Tsai-type 1/1 AAY \AC.
This structure model contains 9 crystallographic sites and includes 174.1 atoms in total in a unit cell, which corresponds to the chemical composition \agyIX.
Note that this composition is compatible with the nominal composition \agyIN. 

Figure~\ref{fig:Crystal} shows the structure model for the \agyI\ \AC. 
As shown in Fig.~\ref{fig:Crystal}(a), the cluster center (M7 site) is occupied by Au atoms with an occupancy of 26.9\%.
The complicated shape of the cluster center, in which 4 atoms in total are ideally included, indicates an average of variously oriented  tetrahedrons.
The second shell [Fig.~\ref{fig:Crystal}(b)] is a dodecahedron, in which there are Au/Ge mixed sites named M2 and M4 sites.
The third shell [Fig.~\ref{fig:Crystal}(c)] is an icosahedron, in which Yb ions exclusively occupy the Yb1 site.
The icosahedron is surrounded by an icosidodecahedron [Fig.~\ref{fig:Crystal}(d)] in which the M1 and M6 sites are almost occupied by Au atoms and the M8 site is occupied by Ge with an occupancy of 95\%.
These shells, i.e., the dodecahedron, icosahedron, and icosidodecahedron compose the so-called Tsai-type cluster.
Finally, Tsai-type clusters are embedded in the cage composed of M3 and M5 [Fig.~\ref{fig:Crystal}(e)]. 
The periodic arrangement of this cage forms a body-centered cubic (bcc) structure, as shown in Fig.~\ref{fig:Crystal}(f). 
A characteristic feature of this model is a chemical ordering of Au and Ge.
These structural properties of \agyI\ \AC\ are very similar to those of Au-Al-Yb \AC. 

The valence electron concentration $e/a$ and the Fermi wave vector $k_{\rm F}$ are calculated to be $1.80$ and $1.17$ nm$^{-1}$, respectively, by assuming valences of Au: 1, Ge: 4, and Yb: 2, where the divalent Yb was deduced from the nonmagnetic properties of the \agyI\ \AC~\cite{Deguchi2014}. 
From a residual resistivity $\rho (0) = 150$ $\mu \Omega$cm, the mean free path is evaluated to be $0.59$ nm.
Note that this value is close to the intercluster distance~\cite{Deguchi2014}. 
This suggests that the large residual resistivity, a common feature of Tsai-type \QCs\ and \ACs , would mainly originate from an orientationally disordered tetrahedron because the M7 site has a large $B$-factor that is proportional to the atomic displacement parameter, $U_{\rm eq} = B/(8\pi^2)$.

The $e/a$ values of the isostructural materials are summarized in Table~\ref{tab:parameter}~\cite{Ishimasa2011,Matsukawa2014,Deguchi2014}.
Typical $e/a$ values for the Tsai-type \QCs\ and \ACs\ are between 1.95 to 2.15.
The \agyI\ \AC\ may also be interpreted as belonging to the Hume-Rothery phase, although the $e/a$ value ($\simeq 1.80$) of the present system is slightly smaller than those of the other ACs.

The values of the lattice parameter $a$ and average atomic radius $\bar{r}$ are also listed in Table~\ref{tab:parameter}. 
Here, $\bar{r}$ was calculated from the radii of divalent Yb ($0.194$ nm) and trivalent Yb ($0.174$ nm)~\cite{PearsonTEXT}.
In Fig.~\ref{fig:AC}, $\bar{r}$ is plotted as a function of $a$ for isostructural Tsai-type ACs.
We observe a linear relationship between them.
This implies that the Yb valence (therefore, magnetism) is determined by the relationship between $\bar{r}$ and $a$. 

\begin{table}[t]
\begin{center}
\caption{Summary of Yb-based Tsai-type 1/1 cubic  \ACs. Yb valence, valence electron concentration $e/a$, lattice parameter $a$, and average atomic radius $\bar{r}$.
}
\label{tab:parameter}
\begin{tabular*}{\columnwidth}[b]{@{\extracolsep{\fill}}lllll}
\hline
Alloys&Yb&$e/a$&$a$ (nm)&$\bar{r}$ (nm)\\
\hline
Au$_{64}$Ge$_{22}$Yb$_{14}$&$+2$&$1.80$&$1.4724(2)$&$0.1496$\\
Ag$_{47}$Ga$_{38}$Yb$_{15}$&$+2$&$1.91$&$1.4687(1)$&$0.1506$\\
Au$_{44}$Ga$_{41}$Yb$_{15}$&$+2$&$1.97$&$1.4527(1)$&$0.1504$\\
Au$_{51}$Al$_{35}$Yb$_{14}$&$+3$&$1.98$&$1.4500(2)$&$0.1480$\\
\hline
\end{tabular*}
\end{center}
\end{table}


\section*{Acknowledgments}
This work was partially supported by Grants-in-Aid for Scientific Research from JSPS, KAKENHI (Nos. 24654102, 25610094, and 26610100). K.D. also thanks the Yamada Science Foundation for financial support.

\bibliography{25321} 


%
\end{document}